\documentclass[12pt]{article} 
\usepackage[letterpaper, margin=1in]{geometry}
\usepackage{url}
\newcommand{\LP}{$\lambda$Prolog} 
\title{A Scheme for Dynamically Integrating C\\ Library Functions into
  a \LP{} Implementation} 
\author{Duanyang Jing} 
\date{May 21, 2019}

\begin{document}
\maketitle

\begin{abstract} 
\noindent The Teyjus system realizes the higher-order logic
programming language \LP{} by compiling programs into bytecode for an
abstract machine and executing this translated form using a simulator
for the machine. 
Teyjus supports a number of builtin relations that are realized
through C code.  
In the current scheme, these relations are realized by including the
C programs that implement them within the simulator and tailoring the
compiler to produce instructions to invoke such code. 
There are two drawbacks to such an approach. 
First, the entire collection of library functions must be included
within the system, thereby leading to a larger than necessary memory
footprint. 
Second, enhancing the collection of builtin predicates requires
changing parts of the simulator and compiler, a task whose
accomplishment requires specific knowledge of these two subsystems. 
This project addresses these problems in three steps.
First, the code for the builtin functions is moved from the simulator
into a library from where relevant parts, determined by information in
the bytecode file, are linked into the runtime system at load time. 
Second, information is associated with each library function about how
it can be invoked from a \LP{} program and where the C code for it is
to be found.
Finally, the compiler is modified to use the preceding
information to include relevant linking instructions in the bytecode
file and to translate invocations to builtin relations into a special 
instruction that calls the dynamically linked code. 
More generally, these ideas are capable of supporting an interface in 
\LP{} to ``foreign functions'' implemented in C, a possibility that is
also discussed.

% \LP{}, a higher-order logic programming language, provides a number of functions
% and relations that are built into the language. As an efficient implementation
% of \LP{}, Teyjus treats these builtins as C procedures implemented in the
% run-time system. This approach requires the code of all builtins being loaded
% together with other components of the run-time system even if no builtins are
% used in the source program, resulting in a larger memory footprint. Moreover, it
% can be difficult to extend builtins by incorporating existing C code, because
% such extension requires manipulating data structures and states internal to the
% run-time system. In this paper we describe a new model for integrating C code
% into the run-time system. It allows a \LP{} programmer to acquire a library
% signature as an interface to a C source program. The interface declares the
% predicates corresponding to functions in the C source program. Then the compiler
% collects information about the external predicates and properly put them into
% the generated bytecode, which is interpreted by the simulator to load the
% library, locate the function and execute the function call. This new model
% reduces the memory footprint of the run-time system because previous builtins
% can now be loaded on the fly only when they are used. It also provides a more
% flexible and easy-to-use interface for programmers to add new C library
% functions as needed. We also describe the potential to extend this model to
% support any C programs as foreign functions in \LP{}.
\end{abstract}

\section{Introduction}\label{sec:intro}

\LP{} is a logic programming language that extends Prolog along several
directions.
The logical fundation of Prolog --- Horn Clauses --- are enriched to
incorporate possiblities of quantifying over function and predicate variables,
and of explicitly representing binding in terms. 
The resulting richer class of
formulas is called higher-order hereditary Harrop formulas~\cite{miller91apal}.
At the programming level, additional features are
implemented such as modular programming, abstract datatypes, higher-order
programming, and the lambda-tree syntax approach to the treatment of bound
variables in syntax \cite{miller12proghol}.

Teyjus is an implementation of the \LP{} language that efficiently addresses
many implementation challenges posed by the new features. 
Underlying the Teyjus system is an abstract machine, referred to
herein as the lpWAM.
This abstract machine inherits from the Warren abstract machine (WAM)
for Prolog \cite{aitkaci91wam} a basic structure for treating  
the unification and search operations that are intrinsic to logic
programming but also incorporates many new mechanisms for treating the
features that are unique to \LP{}.
%
% LpWAM is designed to incorporate
% features intrinsic to 
% logic programming in general, such as backtracking and unification, and also
% features unique to the \LP{} language. 
%
The efficient implementation together with the strong logical
foundations provide us with a powerful logic programming framework.

Not all computations fit natually in the scheme of logic
programming. 
Some operations are entirely side effects, like I/O
operations. 
They lie outside the realm of pure logic, but are still
indispensable for any practical programming languages. 
Other computations can be described in a purely logical manner but
such descriptions can lead to rather inefficient computations. 
For example, arithmetic operations such as addition and division can
be evaluated efficiently using underlying hardware support, whereas
realizing them through a logical description can be quite costly in
time and space. 
As with other languages, \LP{} solves this problem by isolating such
computations in builtin predicates whose concrete implementation can
be provided by means other than logical descriptions. 

The Teyjus system exploits this structure by implementing builtin
predicates via C code. 
Its approach to doing this is based on incorporating the
implementation of these predicates directly into the emulator for
the lpWAM.
There are two problems with this approach. 
First, it requires the code for \emph{all} the builtin predicates to
be included in the runtime footprint, regardless of whether or not the
user program needs them. 
Second, integrating builtins in this way requires knowledge of
relevant parts of the overall implementation, thereby making the task
of adding new builtins more complicated.

We propose an alternative approach to implementing builtins in
Teyjus. 
This approach requires a library builder to construct and maintain an
independent collection of C code and information for using such code
from \LP{} programs. 
The programmer annotates her program based on the latter information.
The compiler then generates instructions for dynamically
linking in relevant parts of the library code and for invoking such
code at the appropriate places. 

The rest of this report explains the problem that we have considered
and our solution to it in more detail. 
In the next section, we provide more specific information about the
structure of the Teyjus system and its treatment of builtins.
In Section~\ref{sec:scheme}, we outline an alternative scheme for
realizing these builtins. 
Sections~\ref{sec:impl} and \ref{sec:eval} then describe the
implementation and evaluation of the new scheme. 
Section~\ref{sec:future} concludes this report by discussing
extensions to the work it presents.

\section{Builtins in the Current Teyjus
  system}\label{sec:teyjus-overview} 

As we have noted earlier, the Teyjus implementation of \LP{} is based
on using an abstract machine: \LP{} programs are translated into code
for the lpWAM, which can then be executed to produce the desired
effects. 
Concretely, four subsystems are used to produce this behavior: a
compiler, a linker, a loader and a simulator for the lpWAM that is
written in C.
The Teyjus system permits \LP{} programs to be organized into a
collection of interacting modules. 
The compiler examines each module, checks its internal consistency and
ensures that it satisfies the promise determined by an associated
signature \cite{teyjus.website}.
If all this checks out, the compiler produces a bytecode file for the
module.
This bytecode file comprises two parts: metadata and instructions to
be executed on the abstract machine. 
The metadata is used by the linker to combine the bytecode files for
the different modules constituting the program into one.
The metadata also provides information to the loader for creating an
initial state for executing the lpWAM instructions. 
Once the loader has done its job, the simulator is ready to accept
user queries and to carry out the computations these queries entail. 

% Teyjus involves four subsystems - a compiler, a linker, a loader and a
% simulator. 
% A \LP{} program may contain several modules. During compilation, a
% \LP{} module is exmained by the compiler to certify the internal consistency and
% to ensure that it satisfies the promies determined by an associated signature
% \cite{teyjus.website}. Then the compiler translates the source program into bytecode
% that contains metadata about the module and also instructions to be executed on
% the abstract machine simulator. The linker takes the bytecode of different
% modules, resolves symbols and absolute addresses and merge instructions as well
% as other metadata segments into one single bytecode file. The loader then reads
% the bytecode file and properly initialize the memory image of the simulator,
% which executes the instructions and communicates the result through an
% interactive user interface.

We are interested in understanding how builtins are treated within
this implementation framework.
The Teyjus manual identifies a collection of predefined predicates
together with their types. 
Programmers can use these predicates directly in their code adhering to
their type declarations, assuming that they denote relations that
follow the semantics that is described for them in the manual. 
These predicates are in fact implemented via routines embedded in the
simulator. 
The use of these predicates in \LP{} programs must eventually
translate into invocations of these C routines. 
To realize this objective, the compiler inserts an instruction of the
form \verb+builtin i+ where \verb+i+ is a numeric index into a
dispatch table that contains a pointer to the appropriate C routine;
executing this instruction will result in a lookup and a transfer of
control to the corresponding procedure. 
Note that the procedure itself will have to convert the arguments,
which have the form of \LP{} data, into a representation suitable for
C and, conversely, it would have to transform the results obtained
from the computation in C into the corresponding \LP{} form; we refer
to these phases as the marshalling and unmarshalling of the data. 
These conversions will typically require access into the data space
of the simulator. 
This is easily realized because the routines are in fact a part of the
simulator.

While the treatment of builtins that we have described is quite simple
and logical, it has two drawbacks. 
First, the code for all the builtins are explicitly integrated into
the simulator, no matter which is actually used in the source
program. 
Depending on how large the collection of builtins is, this could
result in a memory footprint for the runtime system that is
significantly larger than what is actually needed. 
Second, a significant amount of coordination is needed 
between the simulator and the compiler and all this must be manifest
in the code realizing the overall system. 
As a prime example of this, observe that the builtin index used by the
compiler for a predefined predicate must match exactly with what is
contained in the dispatch table. 
Some of this kind of coordination can be realized through
code in the compiler and the simulator that is generated automatically
from a common specification file. 
However, there is still a need to understand and to edit the
compiler and the simulator code in relevant places to add new
builtins. 
This requirement can be daunting to a \LP{} user who wants to provide
new library functions that are realized via C code and thus poses a
barrier to the further development of the system library.

% \subsection{Problems of the current model}
% There are several issues with the above scheme. Firstly all the builtins are
% explicitly integrated into the simulator, no matter which is actually used in
% the source program. This results in a larger memory footprint of the runtime
% system than what is actually needed. Secondly there is no flexible way to extend
% builtins, or more generally, adding external C functions. Adding new builtins
% requires modification of the global builtin table shared across the system, and
% the function dispatch table inside the simulator. Furthurmore, C implementations
% of the builtins involve access to many abstract machine internal data structures
% in order to realize lpwam data representation and calling convention. All these
% knowledge goes beyond the requirement for a C programmer. A more flexible
% interface should allow additional C functions to be integrated to the simulator
% with mimimal requirement of knowledge about system internals, while reducing the
% unnecessary memory footprint to integrate everything in the simulator.

\section{A New Approach to Realizing Builtin Predicates}\label{sec:scheme}

We propose a modified approach to realizing builtin predicates that overcomes
the deficiencies discussed in the previous section.
The central new idea in this approach is to move the code that
implements the builtins out of the simulator and into an external
library from where parts of it can be linked into the user program as
needed.
In this way, only the builtins that are being used are loaded into
memory.
The decoupling also makes the C library independent of the simulator
and the compiler, thereby making it easier for a developer to add new
functionality to the library. 

We need to solve three problems to make this idea work.
First, we have to provide a means for a \LP{} programmer to be able
to use functionality provided by the library without explicit
coordination between the compiler and the simulator.
Second, we have to provide a means for a library developer to build
new functionality---which might have to access the simulator data
spaces at least for the marshalling and unmarshalling
aspects---without having to delve deeply into the simulator code.
Finally, we have to support the possibility of including the needed
parts of library code into the runtime image of an \LP{} program
during the linking and loading phase.
We discuss solutions to each of these problems below.

\subsection{A \LP{} Interface to Library Functionality}

Two items of information are needed to support the use of builtin
predicates in \LP{} programs.
First, the names and types of the predicates that are so defined would
need to be known so that the programmer may use them in a manner that
the compiler can check their uses.
Second, the location of the library and the specific code associated
with a predefined predicate should be known so that a compiler can
generate the necessary linking and dispatch code for the predicate.

The kind of information that is needed can be provided by extending
the modules system already present in Teyjus to realize an interface
to library components.
Typically, a component of the library would consist of code that
implements a collection of builtin predicates.
To enable the use of these predicates, a \LP{} signature file can be
associated with each such component.
This file would provide the location of the component, the types of
the predicates realized, and, for each predicate, a name that
identifies the entry point to the code to be invoked.
Note that, unlike the case for predicates implemented via \LP{} code,
there would be no \LP{} module file corresponding to such a signature
file.
Rather, the identified predicates would be realized by the C code that
would be invoked through special instructions based on the information
in the signature file.
The compiler would need to know to do this, but this matter is easily
handled by enhancing the modules language to suitably distinguish the
inclusion of a builtin ``module'' from that of a vanilla \LP{} module.

\subsection{The Simulator Interface for Library Code}

As noted already, a library developer should be able to write and
compile the C source code for builtin (or external) predicates largely
independently of the simulator. 
The main bottleneck to meeting this requirement is the need to
communicate through shared spaces for both the marshalling and
unmarshalling aspects as well as to ensure conformity with calling
conventions.

To realize the needed independence under the mentioned constraints, a
C header file can be created that includes everything a external C function
might need to access from the simulator.
Library developers would need to assimilate the contents of this
header file to understand how to coordinate their code with the
simulator and also to use any needed functionality already present in
the simulator.
This header file must be included in the file that constitutes the C
source code for the library predicates.
Doing so would allow such a file to be compiled separately and to be
stored in object code form in the library.
This file will also serve as the interface between external C
libraries and the run-time system. 

%% \subsection{How to build the library}
%% Ideally one would like to generate a usable \LP{} library from C source code
%% without much knowledge about the internal implementation of the simulator. The
%% extra work required in a C function is to conform to both the data
%% representations and calling conventions in lpWAM, which requires access to some
%% data structures and global states internal to the run-time system. To realize
%% this a C header file is created that includes everything a external C function
%% might need to access. This file serves as the interface between external C
%% libraries and the run-time system. It should be included in the C library source
%% code to allow the access to the run-time system from external libraries. The C
%% source code should then be augmented to include the marshalling and
%% unmarshalling process. The augmented C code becomes the entry point of the call
%% to library functions. In the end the C source code is compiled to a shared
%% library, which is later loaded into memory and linked with the simulator when
%% the run-time system starts.

%% It is clear that a programmer only needs to understand the interface file in
%% order to build a C library. No other knowledge about the internals of the
%% run-time system is required. In this way the new model is more flexible and easy
%% to use.

\subsection{The Dynamic Linking of Library Components}

The signature files associated with relevant library components
provide information about any additional object code that must be
linked with the simulator to run a given \LP{} program.
The compiler can generate and emit metadata to the bytecode file that
identifies such code.
Most operating systems provide system calls to dynamically load and link a
shared library.
For instance, Unix systems support the \verb+dlopen+ system function that
dynamically loads a shared library and links it with the main program.
The Teyjus loader can be modified to process the additional metadata
to produce a runtime image that includes code realizing library
functionality.

A separate issue that also needs resolution is that of realizing the
appropriate dispatch corresponding to the invocation of builtin
predicates in \LP{} programs.
Operating systems that permit the dynamic loading and linking of code
must also provide a means for resolving symbol references in such
code; for example, such resolution can be realized in Unix systems by
using the \verb+dlsym+ function.
Once a symbol has been resolved, it is an easy matter to realize the
needed dispatch.

An important issue to address is when such symbol resolution should
take place.
It could be done each time an externally implemented predicate is to be
invoked.
However, if the same predicate is invoked multiple times in a \LP{}
program, this can become a costly operation.
A better alternative is to carry out the resolution once at the time
of loading a bytecode file corresponding to a \LP{} program and to
use the absolute address obtained through this process directly in the
dispatch instructions.
This approach is easy to implement.
The compiler can generate a metadata block that lists the names of the
external predicates used in the program and the dispatch instructions
it generates can index into this block.
The loader would then determine the absolute address for each of the
externally implemented predicates, store these addresses in an array and
eventually use this array to change the dispatch instructions into
ones that use absolute addresses as they are loaded into the code space
for the simulator.

\section{An Implementation of the New Scheme}\label{sec:impl}

We have implemented the new approach to supporting builtin predicates
by making various modifications to the existing Teyjus system.
At the outset, we had to make changes to the modules language and to
the instructions to be used to compile invocations of builtin
predicates. 
We then had to make changes to the compiler, the linker, the loader and the
simulator. We describe all these modifications in more detail in
subsections below.

%% To realize the scheme, all the components of the Teyjus system needs
%% to be modified, including the compiler, the linker, the loader and the
%% simulator. Moreover, since these components communicate with each other through
%% the bytecode, additional segments and instructions need to be introduced to
%% convey the information about external functions.

\subsection{Additions to the Modules Language}\label{subsec:modlang}

As discussed earlier, we have to add to the modules language the
capability of describing predicates that are supported through
externally provided C code.
The chief change towards this end is the inclusion in the language of
a new kind of signature file that has the following form:
\begin{verbatim}
sig <signame>.
#lib <clibname>.

extern type <lpname1> <cname1> <type1>.
   ...
extern type <lpnameN> <cnameN> <typeN>.

regcl <lpnames>.
\end{verbatim}
The first line in this declaration sequence associates the name
\verb+signame+ with this signature; this name can be employed in other 
user-defined modules to include the definitions in this signature.
The second line indicates the location of the C library code that
implements the builtin functions that need to be imported into the
\LP{} environment.
This is followed by a sequence of declarations that identify a
collection of builtin predicates together with their types and entry
points that are ostensibly supported by the C library code: in a
declaration of the form 
\begin{verbatim}
extern type <lpname> <cname> <type>
\end{verbatim}
\verb+<lpname>+ stands for the name of the predicate that can be used
in \LP{} programs, \verb+<cname>+ is a symbol that identifies the entry
point and \verb+<type>+ provides the type of the predicate.
The last line in the signature file identifies the subcollection of
the predicates provided in this signature that are ``register
clobbering.'' 
The lpWAM uses a common set of registers for local computations
in a predicate invocation as well as for passing parameters to
predicates. 
Given this, it is generally the case that an invocation of a predicate
can destroy the values that were passed as arguments and hence the
caller has to assume that these contents will not be preserved over
the invocation.  
However, it is useful to know of particular situations in which a
predicate will not destroy the values in argument registers: this
information can be utilized in allocating registers in a way that
minimizes data movement. 
Now, builtin predicates often do not modify argument register values
and hence the compiler uses this as a default assumption.
The last line in the signature file informs the compiler of those
predicates for which it is not safe to make this assumption. 

Given a signature file of this kind, the predicates declared by it can
be used in the code in programmer defined modules by ``accumulating''
the signature.
The compiler must carry out different actions when it is accumulating
code that is obtained from a C-based library from what it would have
to do when accumulating user-defined code.
In light of this, we introduce the new declaration 
\begin{verbatim}
accum_extern <signame>.
\end{verbatim}
to support the accumulation of code from a C-based library.

\subsection{Modification to the Instruction Set}

The existing version of Teyjus has two instructions for invoking
builtin predicates: \verb+builtin x+ and \verb+call_builtin x+ where
\verb+x+ is an index into the builtin dispatch table in the
simulator.\footnote{There are \emph{two} instructions rather than just
  one to accommodate for last-call optimization. The details
  concerning when to use one or the other of these instructions are
  orthogonal to the focus of this project so we do not discuss the
  matter further here.} 
In the new model, the invocation of code for builtin predicates does
not happen through a dispatch table but, rather, by transferring
control to an absolute address.  
Thus these instructions have to be modified to accommodate the new
reality. 

In keeping with the above observation, we have complemented the
mentioned instructions with two new instructions \verb+execute_extern x+ and 
\verb+call_extern x+ in which \verb+x+ represents an absolute address. 
When the compiler generates these instructions, it tentatively uses an
index into an array that will be built at load time from metadata in
the bytecode file and that will be filled in with absolute addresses
determined for the relevant builtins. 
The loader will eventually replace the indices in the compiled versions
of these instructions with absolute addresses to make them ready for
execution. 

\subsection{Modification to the Compiler}

There are two broad changes that have to be made to the compiler to
accommodate the new treatment of builtins: 
\begin{itemize}
\item It must be extended to treat the inclusion of builtin predicates
  in user programs through the new kind of signature files.

\item It must use the new instructions to compile invocations of
  builtin predicates in the user code. 
\end{itemize}

We have added code to the compiler to process signature files
for predefined predicates. 
As with signature files for modules implemented via \LP{} definitions,
this processing adds predicate names together with types associated
with them to the symbol table; this step ensures that the necessary
information for checking the usage of these predefined predicates in
user code is in place.
Entries for these predicates in the symbol table are marked in a way
that indicates that a different kind of instruction must be generated
to realize their invocation. 
In contrast to the accumulation of regular \LP{} modules, the compiler
\emph{does not} look for code realizing the builtin predicates.
Instead, it collects the library and entry point names for
each predefined predicate and maintains them in a list to be used
later in the compilation process. 

The bytecode generation phase incorporates two changes. 
First, metadata is emitted for use by the linker and loader to realize
the process of dynamically linking in the C code that defines the
builtin predicates. 
The information for producing this metadata is available from the
analysis of the signature file as we have just noted. 
Second, the new form of instructions must be generated for the
invocation of builtin predicates. 
It should be evident from all that has been said that this step is
also easy to realize.

\subsection{Modification to the linker}

The linking process produces one single bytecode file from separate bytecode
files compiled from different modules. 
Focusing only on the treatment of predefined predicates, this means
that the metadata segments for such predicates that come from
different bytecode files must be combined into one consolidated
segment. 
The main complexity in doing this is that the indices in the
instructions that invoke the predefined predicates must also be
relativized to the metadata segment that is so generated. 
However, even this is not difficult to do: if the consolidated
metadata segment is obtained via a linear combination of the
individual segments, then the indices for the calling instructions in
the separate bytecode files need only be adjusted by a fixed offset. 

% \subsection{Modification to the linker}
% The linking process produces one single bytecode file from separate bytecode
% files compiled from different modules. After compilation the information from
% individual bytecode files are local to the module itself, such as indices of
% symbols and relative addresses. Now to facilitate external functions one
% additional segment is added to the bytecode, which is the external function
% table. The instructions ``call extern x`` and ``execute extern x`` all have
% operands refering to index in this table. To produce a meaningful single
% bytecode file, the linker needs to merge the external function table segment
% from different modules and update the operands for all ``extern`` instructions.
% This process is trivial to realize. The linker just scans the external function
% table from each individual bytecode and put all the entries together in the
% final linked bytecode file.

\subsection{Modification to the loader}
The loader reads the bytecode file produced by the linker and sets up the memory
image of the run-time system. 
In the new model, the loader should also load the external libraries to
be used, and resolve symbols to libraries to absolute 
addresses. This can be realized by reading the external function table segment and
making the dynamic linking system call for every entry in the table. Later when
the code region is being loaded, the loader can examine every instruction and
update the operands of extern instructions to be the absolute address.

\subsection{Modification to the simulator}
The work required of the simulator with respect to the treatment of
builtins is minimal.
All calls to external functions have already been resolved to absolute 
addresses by the loader, 
and the functions are augmented to properly manipulate data representations and
some states of the simulator. As a result the simulator just needs to invoke the
functions. 
%GN Why are you repeating this?
% This is much less work for the simulator compared to the old model,
% where the simulator keeps track of a builtin dispatch table that stores an array
% of function pointers to the implementation of builtins. This part can be almost
% completely removed from the simulator, with only few builtins as
% exceptions.
Some builtins are instrinsic to the \LP{} language and it makes
better sense to leave the code realizing them within the simulator
system. These builtins include 
``solve'', ``not'', ``eval'' and comparison predicates.
As a result, the builtin dispatch table and the old instructions for
invoking builtin predicates and completely removed from the system. 
However, only a small number of builtin predicates are treated in the
old way and hence there is only a marginal overhead in the footprint
in the case that these predicates are not actually employed in the
user code.

\section{Building Libraries Using the New Scheme}\label{sec:eval}

%\section{Evalution of the New Scheme}\label{sec:eval}

%\textbf{GN: This section is not really coherent and needs to be
%  revised. To begin with, the first part, that talks about overheads does not
%  fit in with the second and also is not really providing any real
%  information. Further, the second part is not about ``evaluation'' in
%  any sense. I have changed the title of the section to reflect this
%  part of the content better, but the content is still not
%  sensible; it needs to focus on how one builds a library rather than
%  discussing specifics of marshalling and unmarshalling for example.}

\noindent
In this section a more practical view of the proposed scheme is discussed, in
particular how to extend C code to produce a C-based library, which realizes
the original C computation in conformity with the data representation and
calling conventions of the Teyjus simulator.
As discussed earlier, a library developer needs to deliver two files as
an external library.
One is the \LP{} signature file that describes the predicates implemented
in the library, the other is the C implementation compiled as a shared
library.
Suppose a library developer wants to provide a library with various
mathematical predicates, much like the C header file math.h:
\begin{verbatim}
sin: real -> real -> o.
cos: real -> real -> o.
tan: real -> real -> o.
  ...
\end{verbatim}
We will discuss in detail how this can be realized in the new scheme.

The starting point would be to acquire the C implementation of these functions.
In this example the library developer can simply use the implementation 
provided in the header file math.h, or write their own implementation.
Then the library developer would need to extend the C code to make it
conform to the data representation and calling convention of lpWAM.
The simulator interface for library code is designed specifically for this
purpose, which is a C header file that encapsulates some parts of the simulator 
that might be called by library code.
In this case two functions would be useful:
\begin{verbatim}
float TJ_getReal(int i);
void TJ_returnReal(int i, float val);
\end{verbatim}
The first function takes the data at argument register i and converts it to a
C float representation.
The second function takes a C float, converts it into a real term in lpWAM,
and unifies it with the existing data at argument register i.
Conceptually the first function should be called on input arguments, and the
second function should be called on the argument register that holds the term 
to be unified with the actual return value.
Now the library developer can write a wrapper function
that wraps the original C function using the above two functions:
\begin{verbatim}
#include <math.h>

void sin_wrapper()
{
    float a1 = TJ_getReal(1);
    double ret = sin((float)a1);
    TJ_returnReal(2, (float)ret);
}
\end{verbatim}
This wrapper function can serve as the entry point from the simulator to
the library code, which includes all the work during the invocation of the
\verb|sin| predicate.
Because the "return" function in the end transfers control to some other
procedures in the simulator, the wrapper function should have \verb|void| type
and take no arguments.
Other functions would follow the same pattern because of their similarity in 
nature.
For predicates with different arity and types of arguments, the library
developer just needs to find other functions and call them accordingly
in the wrapper function.
In the end the library developer would compile the C file that includes all 
the wrapper functions into a shared object file and deliver it to \LP{} programmers.

The \LP{} signature file is another file the library developer needs to
produce. 
The syntax for this file is discussed in Section~\ref{subsec:modlang}. 
With the C code extended and compiled, the library developer would now have all
information required for this signature file. 
As an example, suppose the compiled shared object file has name \verb|math.so|, 
the signature file might have the following structure for the 
mathematical predicates:
\begin{verbatim}
#sig math
#lib math

extern type sin sin_wrapper real -> real -> o.
extern type cos cos_wrapper real -> real -> o.
extern type tan tan_wrapper real -> real -> o.
  ...
#regcl sin
\end{verbatim}
Section ~\ref{subsec:modlang} also discussed the need to
inform the compiler of those predicates that modify argument register values
during invocation. 
Some functions in the simulator interface would have this
behavior. 
Calling these functions in the extended C code would require the library
developer to include the predicate in the list of \verb|#regcl| predicates.

\section{Some Ideas for Further Developments}\label{sec:future}
In this report we have presented a scheme that allows \LP{} builtins
that are implemented through C code to be treated as components that
are loosely coupled with the main Teyjus system.
This scheme has the virtue of requiring a library developer to have
only a rudimentary understanding of the simulator---as presented
through the interface declarations---to extend existing C code to
external libraries. 
In reality, this scheme has the potential for providing a systematic
way to integrate any external C computation into the Teyjus system.
It would be of interest to explore its development in this direction.

The main issue to deal with in realizing this kind of generalization
is to determine how the simulator interface can be enriched to 
provide a means for marshalling and unmarshalling between complex \LP{} data
encodings and corresponding C data structures.
The current simulator interface only supports primitive types in Teyjus, including 
\verb|int|, \verb|real|, and \verb|string|, which are merely the starting point 
of the Teyjus type system. 
Structured data types introduced by \verb|kind| declarations and data
constructors introduced by \verb|type| declarations are not supported.
%
%% In this section we discuss the possibility to further expand the scheme so
%% that it can support structured data in Teyjus. 

To add support for structured data, we would need to compare typical 
data encodings in \LP{} with the approach used in C. 
As a typical example, suppose a library developer would like to provide predicates 
on a pair object:
\begin{verbatim}
kind pair type -> type -> type.
type pr   int -> int -> pair int int.
\end{verbatim}
Such an object might be represented in C by using a structure that has
two integer fields: 
\begin{verbatim}
struct pair {  int x;  int y; };
\end{verbatim}
Once the data representations that one needs to translate between are
known, it should be easy to write the code to effect such
translations.
In fact, such marshalling and unmarshalling code can even be generated
automatically.
Thus, the key issue is how the relevant knowledge about the relevant
data representations may be communicated to the library developer.
%% Unlike the primitive types, the structure of these types are arbitrarily defined by 
%% programmers, hence there is no way to hard code the marshalling and unmarshalling 
%% procedures in the simulator interface file.
One possible approach to doing this is to use a specification file
that describes both the \LP{} data objects and the corresponding C
representation.
Thus, the key task in this endeavor becomes that of describing a
convenient and flexible structure for such a specification file. 

Once a format for writing such specifications has been described, the
next interesting task would be to develop a ``marshalling and
unmarshalling code generator'' component of the Teyjus system that
would read the specification file, and automatically generate the code
for going between the \LP{} and C representations. 
In fact, this translation generator component can be treated as one additional 
layer of abstraction between the library developer and external libraries,
which could take care of not only the generation of all marshalling and 
unmarshalling code, but also the generation of the \LP{} signature file.
In the end, a library developer would only need to provide a specification file 
that contains high level descriptions of the \LP{} data objects and
the corresponding C data structures, the \LP{} predicates, and the
C code that is to realize the functionality and then the library generator would 
automatically produce the shared library object as well as the \LP{} signature 
file. 

The development of both the specification format and
component for generating the translation code seems to be a natural
next step to this project.
Moreover, carrying it out holds the exciting promise of transforming the
scheme we have described in this report into a more general and
flexible interface between \LP{} and C programs. 
Unfortunately, playing these ideas out in detail is beyond the scope
of this Master's project and has therefore to be left to future work.

\section*{Acknowledgements}

This material is based upon work partially supported by the National
Science Foundation under Grant No. CCF-1617771. Any opinions,
findings, and conclusions or recommendations expressed in this
material are those of the author and do not necessarily reflect the
views of the National Science Foundation.

\bibliographystyle{plain}
\bibliography{master.bib}

% \begin{thebibliography}{9}
% \bibitem{wambook} H ASSAN A ÏT -K ACI.  Warren's Abstract Machine - a tutorial
% reconstruction.  Reprinted from MIT version. February 18, 1999.

% \bibitem{teyjusdesc} Nadathur G., Mitchell D.J. (1999) System Description:
% Teyjus—A Compiler and Abstract Machine Based Implementation of λProlog. In:
% Automated Deduction — CADE-16. CADE 1999. Lecture Notes in Computer Science, vol
% 1632. Springer, Berlin, Heidelberg

% \bibitem{pl-ravi} Ravi Sethi. Programming languages - concepts and constructs,
% second edition. Addison Wesley Longman. 1996

% \bibitem{lpweb} aaa

% \bibitem{uniformproof} Dale Miller, Gopalan Nadathur, Frank Pfenning, and Andre
% Scedrov, "Uniform Proofs as a Foundation for Logic Programming". June 1989.

% \bibitem{teyjusbuiltins} Shyan-Ming Perng, "A Realization of Built-in Procedures
% in a \LP{} Implementation". August 5, 1998
% \end{thebibliography}

\end{document}